\documentclass[aps,prl,twocolumn,superscriptaddress]{revtex4}

\usepackage{amssymb}
\usepackage{graphicx}
\usepackage{amsmath}
\usepackage{braket}

\usepackage{hyperref}

\begin{document}
	
\title{Topological Sachdev-Ye-Kitaev Model}
\author{Pengfei Zhang}
\affiliation{Institute for Advanced Study, Tsinghua University, Beijing, 100084, China}
\author{Hui Zhai}
\affiliation{Institute for Advanced Study, Tsinghua University, Beijing, 100084, China}
\affiliation{Collaborative Innovation Center of Quantum Matter, Beijing, 100084, China}
\date{\today}
	
\begin{abstract}

In this letter we construct a large-N exactly solvable model to study the interplay between interaction and topology, by connecting Sacheve-Ye-Kitaev (SYK) model with constant hopping. The hopping forms a band structure that can exhibit both topological trivial and nontrivial phases. Starting from a topologically trivial insulator with zero Hall conductance, we show that interaction can drive a phase transition to topological nontrivial insulator with quantized non-zero Hall conductance, and a single gapless Dirac fermion emerges when the interaction is fine tuned to the critical point. The finite temperature effect is also considered and we show that the topological phase with stronger interaction is less stable against temperature. Our model provides a concrete example to illustrate interacting topological phases and phase transition, and can shed light on similar problems in physical systems.  

\end{abstract}
	
\maketitle

Sachdev-Ye-Kitaev (SYK) model \cite{Kitaev2,SY} has recently drawn a lot of interests from both condensed matter \cite{Kitaev2,Comments,spectrum1,spectrum2,spectrum3,Liouville,Liouville2,SYK new,SYK new2,SYK new3,SYK new4,quench1,operator growth,Balent,our,Altman,sk jian,yyz condensation,Constant,sentil,thermal transport,Yingfei1,Yingfei2,Kondo,sk2,exp} and gravity physics communities \cite{Comments,bulk Yang,bulk spectrum Polchinski,bulk2,bulk3,bulk4,bulk5,syk-bh,SYK g new1,SYK g new2,SYK g new3,new g,new g2}, because it displays an emergent conformal symmetry, holographic duality to AdS$_2$ gravity and maximally chaotic behavior. In addition to its significant impact on the AdS/CFT research, SYK model is also of great interests from a pure condensed matter physics perspective. Let us use SYK dot to refer to a cluster of $N$ Majoarana fermion modes or complex fermion modes with all-to-all random interactions. The Green's function of such an SYK dot can be exactly solved in the large-N limit. More importantly, this large-N limit is strongly in contrast to many other large-N solvable models in condensed matter systems \cite{Coleman,Fradkin}, where the leading order action is a quadratic one and the resulting quantum state is essentially a free one. The leading order solution of an SYK dot is a strongly correlated state and displays non-Fermi liquid type behavior \cite{Comments}. Hence, the SYK dot can be used as the building blocks for granule construction of exactly solvable strongly correlated models by connecting SYK dots with tunneling \cite{Balent,our,Altman,sk jian,yyz condensation,Constant,sentil,thermal transport,Yingfei1,Yingfei2,Kondo,sk2,exp}. Depending on how to connect them, different types of interacting physics can emerge. These solvable models can be used to shed insight on fundamentally important and open issues in condensed matter physics, such as the non-Fermi liquid states and phase transitions between two non-Fermi liquid phases \cite{our}, as well as from a non-Fermi liquid phase to another phases \cite{Altman,sk jian,yyz condensation,sk2}. 

In this work we consider SYK dots coupled by constant quadratic hopping, and the quadratic hopping itself forms a topological band. Our model, for the first time, combines the physics of SYK interaction with topological band theory to address the interaction effects in topological theory, and is therefore termed as ``topological Sachdev-Ye-Kitaev model". Previously, topological phases with interactions has been classified for both bosons and fermions \cite{cl review,XC scince}, for instance, by using field theory approaches \cite{WZ1,WZ2,WC,WC2,Yuanming}, or by utilizing mathematical tool of group cohomology \cite{XC scince,XC PRB,Gu1,Gu2,Ran}, or by constructing exactly solvable models with specially designed lattice structures \cite{X.C.,SQ Ning,Ran,Gu1,Gu2}. Our topological SYK model provides an alternative class of exactly solvable interacting topological model. It does not depend on any specific lattice structures and can explicitly show that the interaction can drive a transition from a topological trivial phase with vanishing Hall conductance to a topologically nontrivial phase with quantized Hall conductance. 

\begin{figure}[tp]
\includegraphics[width=2.5 in]{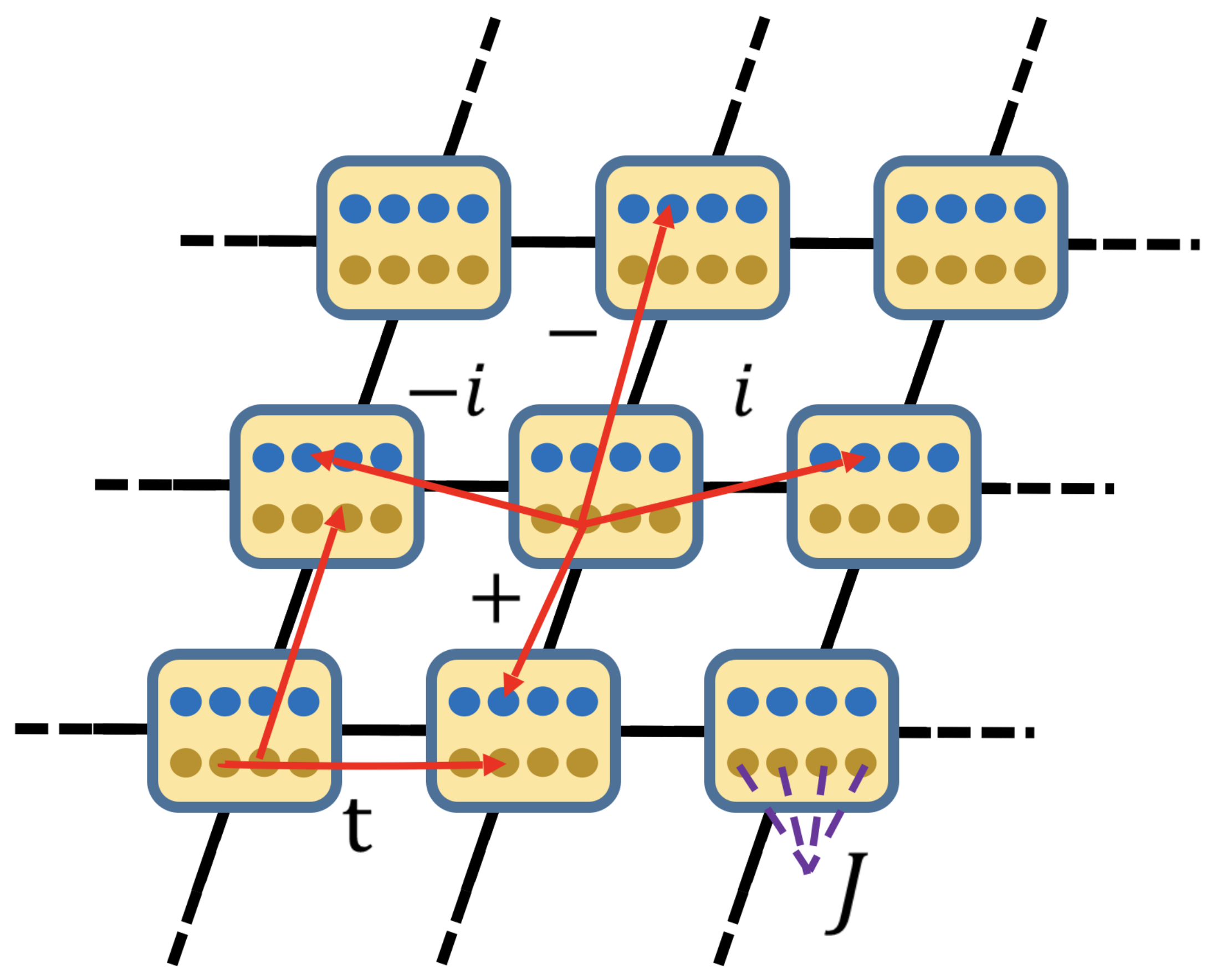}
\caption{Schematic of the topological Sachdev-Ye-Kitaev model. On each site, different color represents different pseudo-spin degree of freedom. For same color, different point denotes $N$-different flavours. Solid lines with arrow denotes quadratic hopping within the same flavours, and the dashed line denotes on-site interaction within the same pseudo-spin.}
\label{Scheme}
\end{figure}

\textit{Model.} Our model is schematically shown in Fig. \ref{Scheme}. We consider complex fermion in a two-dimensional lattice with $i$ being the site (or unit cell) index. Within each site, there is spin or pseudo-spin (such as two sublattices within each unit cell) degree of freedom denoted by $\sigma=\uparrow,\downarrow$ and flavour index denoted by $\lambda=1,\dots,N$. We consider that hopping only takes place between fermions with the same flavour index. Thus, the single-particle Hamiltonian is simply $N$ decoupled copies of two-band Chern insulator model, that is written as
\begin{align}
\hat{H}_0=\sum\limits_{\lambda=1}^{N}\sum\limits_{\mathbf p}\sum\limits_{\sigma\sigma^\prime}\hat{c}^{\dagger}_{\mathbf p\sigma\lambda}h_{\sigma\sigma^\prime}(\mathbf p)\hat{c}_{\mathbf p \sigma^\prime\lambda},\label{H_0}
\end{align}
where ${\mathbf p}$ is the quasi-momentum. Here we consider a concrete form of $h(\mathbf p)$ as
\begin{align}
h(\mathbf p)=&t_0(m-\cos(p_x)-\cos(p_y))\sigma_z\notag\\&+t_0\sin(p_x)\sigma_x+t_0\sin(p_y)\sigma_y.
\end{align} 
For each copy, it has been shown that for $|m|>2$ the Chern number $\mathcal{C}=0$ and the insulator is a topological trivial one, and $\mathcal{C}=\pm 1$ for $0<m<2$ and $-2<m<0$, respectively, which leads to two topological nontrivial insulators. 

Now, following the original proposal of SYK model, we introduce on-site interaction between the $N$-different flavors but with the same spin index, that is
\begin{align}
\hat{H}_\text{int}=\sum\limits_{i\sigma}\sum_{\lambda_1,\lambda_2,\lambda_3,\lambda_4}\frac{J^{i,\sigma}_{\lambda_1\lambda_2\lambda_3\lambda_4}}{4}\hat{c}^{\dagger}_{i\sigma\lambda_1}\hat{c}^{\dagger}_{i \sigma \lambda_2}\hat{c}_{i\sigma\lambda_3}c_{i\sigma\lambda_4},\label{Hint}
\end{align}
where each $J^{i\sigma}_{\lambda_1\lambda_2\lambda_3\lambda_4}$ are independent Gaussian random variables that satisfies 
\begin{align}
&\overline{J^{i\sigma}_{\lambda_1\lambda_2\lambda_3\lambda_4}}=0,\,\,\,\,\,\,\overline{|J^{i\sigma}_{\lambda_1\lambda_2\lambda_3\lambda_4}|^{2}}=\frac{2J^2}{N^3}. \label{Dis J}
\end{align}
This choice of Eq.\eqref{Dis J} ensures a well-defined large-$N$ limit. It is important for later discussion to emphasize that any two $J$-coefficients are two independent Gaussian variables and their correlation after disorder average vanishes, if any one of their total six labels is different. 

The total Hamiltonian is therefore given by 
\begin{equation}
\hat{H}=\hat{H}_0+\hat{H}_\text{int}. \label{H}
\end{equation}   
Here we have set the chemical potential $\mu=0$. This is crucial to ensure that the ground state with $\hat{H}_0$ alone is a gapped insulator and with $\hat{H}_\text{int}$ alone is a non-Fermi liquid phase.  

\textit{Spectral function.} The single particle spectral function of this model is exactly solvable in the large-N limit. As far as the leading order solution in the large-N expansion is concerned, there are a few features that we should emphasize here. 

\begin{figure}[t]
\includegraphics[width=3.4 in]{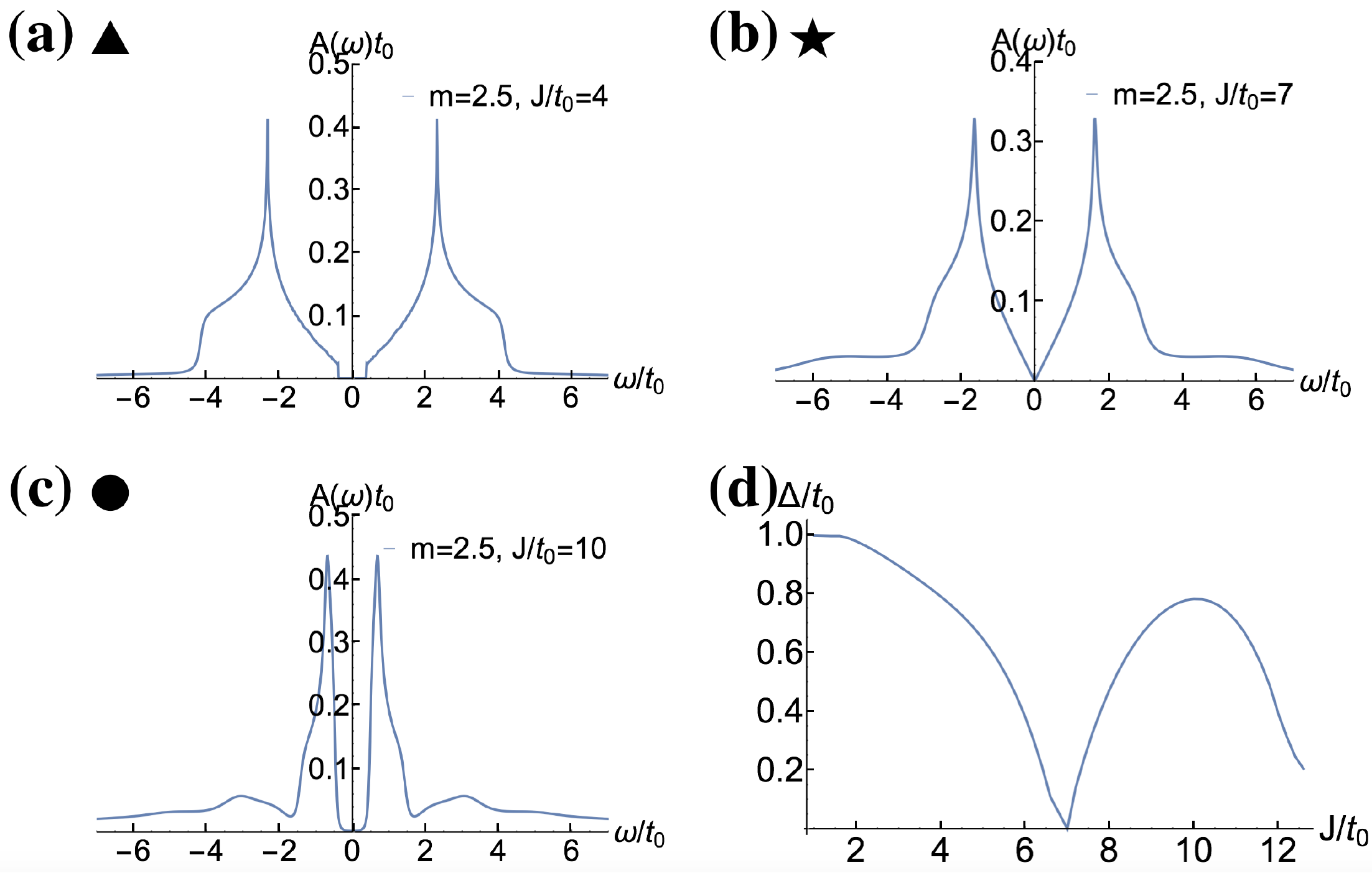}
\caption{(a-c). The spectral functions $A(\omega)$ for $m=2.5$ and $J/t_0=4,7,10$, respectively, located at marked points in the phase diagram Fig. \ref{phasediagram}. (d). The evolution of gap $\Delta$ when $J/t_0$ increases with fixed $m$. }
\label{spectral}
\end{figure}

First of all, the two-point Green's function is diagonal in the flavor space and is independent of the $\lambda$ index, despite that the SYK interaction does mix different different flavors. This is because the disorder average over the $J$-coefficient mentioned above forces the incoming and the outgoing flavor index to be identical. Hence, the later discussion of Chern number and Hall conductance refer to those of each flavor. 

Now we first introduce the two-point Green's function, ignoring the $\lambda$ index, as  
\begin{align}
(G(\tau,\mathbf x))^{\sigma\sigma^\prime}=\left<\mathcal T_\tau \hat{c}^
\dag_{i+\mathbf x,\sigma\lambda}(\tau)\hat{c}_{i,\sigma^\prime\lambda}(0) \right>,
\end{align}
and $G(i \omega_n,\mathbf p)$ is the Fourier transform of $G(\tau,\mathbf x)$, which can be written as 
\begin{align}
G^{-1}(i \omega_n,\mathbf p)&=-i\omega_n+h(\mathbf p)-\Sigma(i\omega_n).\label{Green}
\end{align}
Here both Green's function $G$ and the self-energy $\Sigma$ is a $2\times 2$ matrix written in the spin space. $\Sigma(\tau)$ be the Fourier transfer of $\Sigma(i\omega_n)$.

Secondly, we shall emphasize some properties of the self-energy that is quite useful for later discussion. Also because of the disorder average, the incoming and outgoing legs in the self-energy diagram have to have same spin and position indices, and therefore, (i) $\Sigma(\tau)$ is diagonal in the spin space, and (ii) $\Sigma(\tau)$ is also a pure local one without any momentum dependence \cite{fn}. With these, $\Sigma(\tau)$ can be written as 
\begin{align}
\Sigma(\tau)&=\begin{pmatrix}
\Sigma^{\uparrow}(\tau)&0\\
0&\Sigma^{\downarrow}(\tau)
\end{pmatrix},\label{Sigma}
\end{align}
and following the standard procedure of solving the original SYK model, it can be show that the self-energy satisfies following equation 
\begin{align}
\Sigma^{\sigma}(\tau)=J^2(G^{\sigma\sigma}(\tau))^2G^{\sigma\sigma}(-\tau). \label{self-energy}
\end{align} 
Thus, Eq. \ref{Green} and Eq. \ref{self-energy} form a coupled self-consistency equation for solving the Green's function. 

Thirdly, we show that the spectral function is still generically gapped despite of the SYK interactions. Considering the SYK fixed point at zero tunneling limit, the scaling dimension of the single fermion operator is $1/4$. Now turning on the quadratic hopping, because the scaling dimension of a bilinear operator is then $1/2$, it is a relevant perturbation. Hence, the low-energy behavior at $\omega\rightarrow 0$ is dominated by the quadratic hopping. And because the quadratic hopping generically gives rise to an insulator with a gapped spectral at low-energy, from this point of view we expect the system remains generically gapped even when the interaction is turned on. 

This statement can be verified by numerically solving the self-consistent equation and obtaining the spectral function. This caluclation is most easily done in the real time. We first apply the analytical continuation to the self-consistent equations \eqref{Green} and \eqref{self-energy} to obtain corresponding self-consistent equation for the retarded Green'€™s
function $G_R(\omega,\mathbf p)$ \cite{supple}. By solving these equations numerically, we could determine the spectral function averaged over momentum and spin as 
\begin{align}
A(\omega)=-\frac{1}{2\pi}\int\frac{d^2p}{(2\pi)^2}\text{Im\ Tr\ }G_R(\omega,\mathbf p).
\end{align} 
Here we need to emphasize that particular attention should be paid to the finite size scaling for this numerical calculation \cite{supple}.
The results for $m=2.5$ and $J/t_0=4,7,10$ are shown in FIG. \ref{spectral}(a-c). We can see from Fig. \ref{spectral}(a) and (c) that a clear U-shaped spectrum near zero-frequency, which reveals a gap state and also allows us to identify the value of gap unambiguously. In Fig. \ref{spectral}(d), we show the gap value as a function of interaction strength for a fixed $m$. It is clear that the gap is generically non-zero except at one single point, which we will later identify as the critical point. 

One notices from Fig. \ref{spectral}(d) another interesting feature, that is, for large $J$, the gap also displays a non-monotonic behaviors. This is because the SYK interaction plays a dual role. On one hands, as we will show below, it renormalizes single particle Hamiltonian and drives the topological transition, and the gap shall increase as moving away from the critical point. On the other hand, when the interaction becomes stronger, its effect will become more and more significant once the system is slightly away from the low-energy limit, and since the SYK interaction itself favors a non-Fermi liquid type gapless state, it will eventually make the gap smaller. 

Finally we remark that the spectral function $A(\omega)$ is symmetric under $\omega\rightarrow -\omega$. This is because of a non-local symmetry $\mathcal P$ defined as follows:
\begin{align}
\mathcal{P}c_{\mathbf p\sigma\lambda}\mathcal{P}^{-1}=(\sigma_y)_{\sigma\sigma^\prime}c^{\dagger}_{\mathbf p\sigma^\prime\lambda}, \label{P-symmetry}
\end{align}
which can be understood as a combination of particle-hole transformation and inversion. 

\begin{figure}[t]
\includegraphics[width=3.2 in]{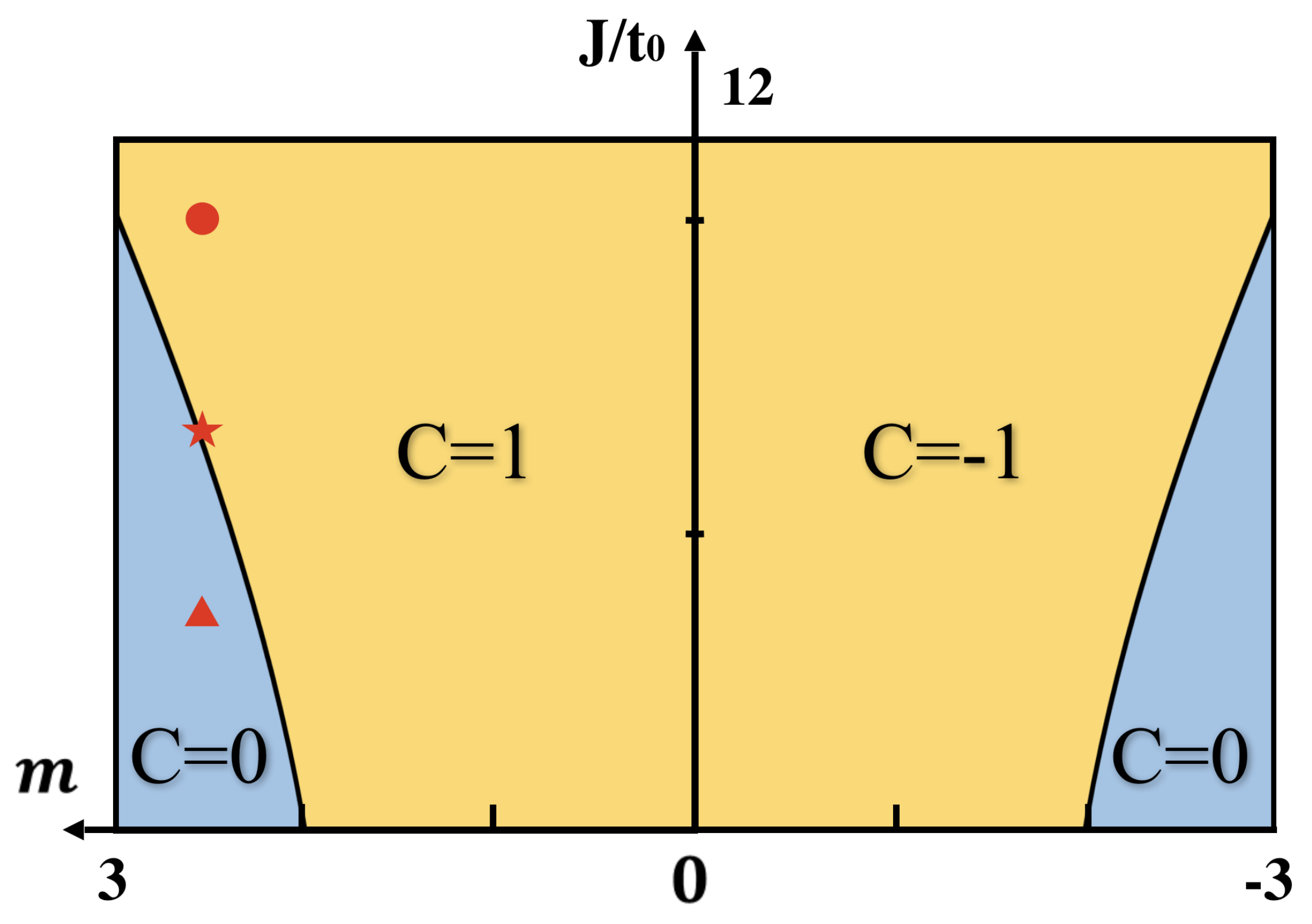}
\caption{The phase diagram of the system in term of single particle mass $m$ and the interaction strength $J/t_0$, where the Chern number is calculated for each flavor. The spectral function of these three marked points are shown in Fig. \ref{spectral}. }
\label{phasediagram}
\end{figure}

\textit{Chern Number and Phase Diagram.} Because the system is always gapped, we can calculate the Chern number of this interacting system with the method introduced in Ref. \cite{WZ1}. Here we define an effective Hamiltonian as 
\begin{equation}
h_{\text{eff}}(\mathbf p)=-G_R^{-1}(\omega=0,\mathbf{p}),
\end{equation}
and let $|\alpha_-(\mathbf p)\rangle$ be the eigenstate of $h_{\text{eff}}(\mathbf p)$ with negative eigenvalue. We can obtain the Chern number by integrating the Berry curvature using the standard formula
\begin{align}
&{\bf A}(\mathbf p)=i\langle\alpha_-(\mathbf p)|\boldsymbol{\nabla}|\alpha_-(\mathbf p)\rangle,\label{def A}\\
&\mathcal{C}=\frac{1}{2\pi}\int d^2p \left( \partial_x A_y(\mathbf p)-\partial_y A_x(\mathbf p) \right).\label{def C}
\end{align}
This gives different values of $C$ equalling zero, or $\pm 1$, and the phase diagram is shown in Fig. \ref{phasediagram}. For non-interacting case $J=0$, the phase diagram is determined by the parameter $m$ alone. And for interacting case, it only depends on one extra parameter that is $J/t_0$. 

In fact, there is an even simple way to determine the phase diagram. It utilizes two facts: (i) The self-energy $\Sigma(\omega)$ is diagonal and is not a function of momentum, as discussed above. (ii) Due to the $\mathcal{P}$-symmetry defined in Eq. \ref{P-symmetry}, it can be proved that $\Sigma^\uparrow(0)=-\Sigma^\downarrow(0)$ \cite{supple}. Thus, $h_{\text{eff}}({\bf p})$ only differs from $h({\bf p})$ by a $\Sigma^\uparrow(0)\sigma_z$ term and it renormalizes the mass from $m$ to 
\begin{equation}
m_{\text{eff}}=m-\frac{\Sigma^\uparrow(0)}{t_0}.
\end{equation}
Thus the phase boundary is simply determined by $m_{\text{eff}}=0,\pm 2$. Moreover, the phase diagram of Fig. \ref{phasediagram} appears to be symmetric with respect to $m \rightarrow -m$, and this is because under an anti-unitary symmetry transformation \begin{align}
\mathcal{T}c_{\mathbf p,\sigma\lambda}\mathcal{T}^{-1}=(\sigma_y)_{\sigma\sigma'}c_{\mathbf{(\pi,\pi)}+\mathbf p,\sigma'\lambda}\ \ \ \ \mathcal{T}i\mathcal{T}^{-1}=-i,
\end{align} 
with the disorder average, the only changes to the Hamiltonian is $m\rightarrow -m$, and this anti-unitary transformation changes $\mathcal C \rightarrow -\mathcal C$. 

The most dramatic feature of the phase diagram is that the interaction actually enlarges the topological nontrivial regime. When $|m|>2$, the non-interacting case is a topological trivial state, and the interaction can actually brings the system into a topological nontrivial regime. This appears quite counter-intuitive. In fact, this can be understood by perturbative calculation in term of interaction strength $J$, which shows that $\Sigma^\uparrow=\alpha m J^2$ with $\alpha>0$ \cite{supple}. This means that, independent of the sign of $m$, the interaction always renormalizes the absolute value of $m$ toward a smaller value, which a topological nontrivial state is favored. 

\begin{figure}[t]
\includegraphics[width=2.4 in]{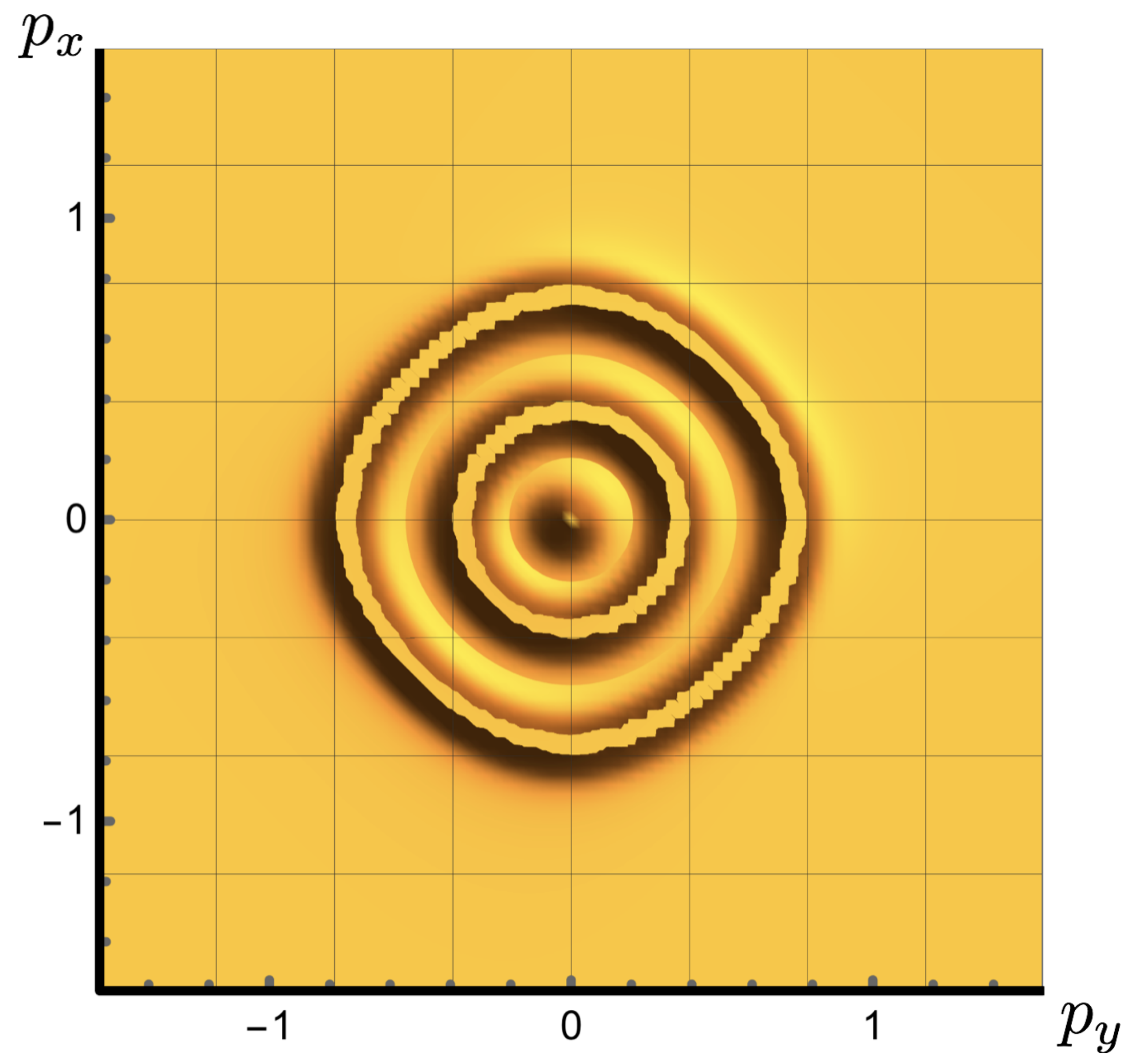}
\caption{The momentum resolved spectral function $A(\omega,{\bf p})$ for $m=2.5$, $J/t_0=7$. Three different contours correspond to peaks of $A(\omega,{\bf p})$ with $\omega/t_0=0$, $0.3$ and $0.6$, respectively. }
\label{Dirac}
\end{figure}

Our model can also clearly illustrate the critical theory. Fig. \ref{spectral}(b) shows that the spectral function displays a V-shape at zero-energy right at the phase boundary. In fact, a momentum resolved plot of the spectral function $A(\omega,{\bf p})$ can clearly show a single Dirac point in the momentum space, as shown in Fig. \ref{Dirac}. Therefore, we believe the critical theory is a free Dirac theory. 

\begin{figure}[t]
\includegraphics[width=2.8 in]{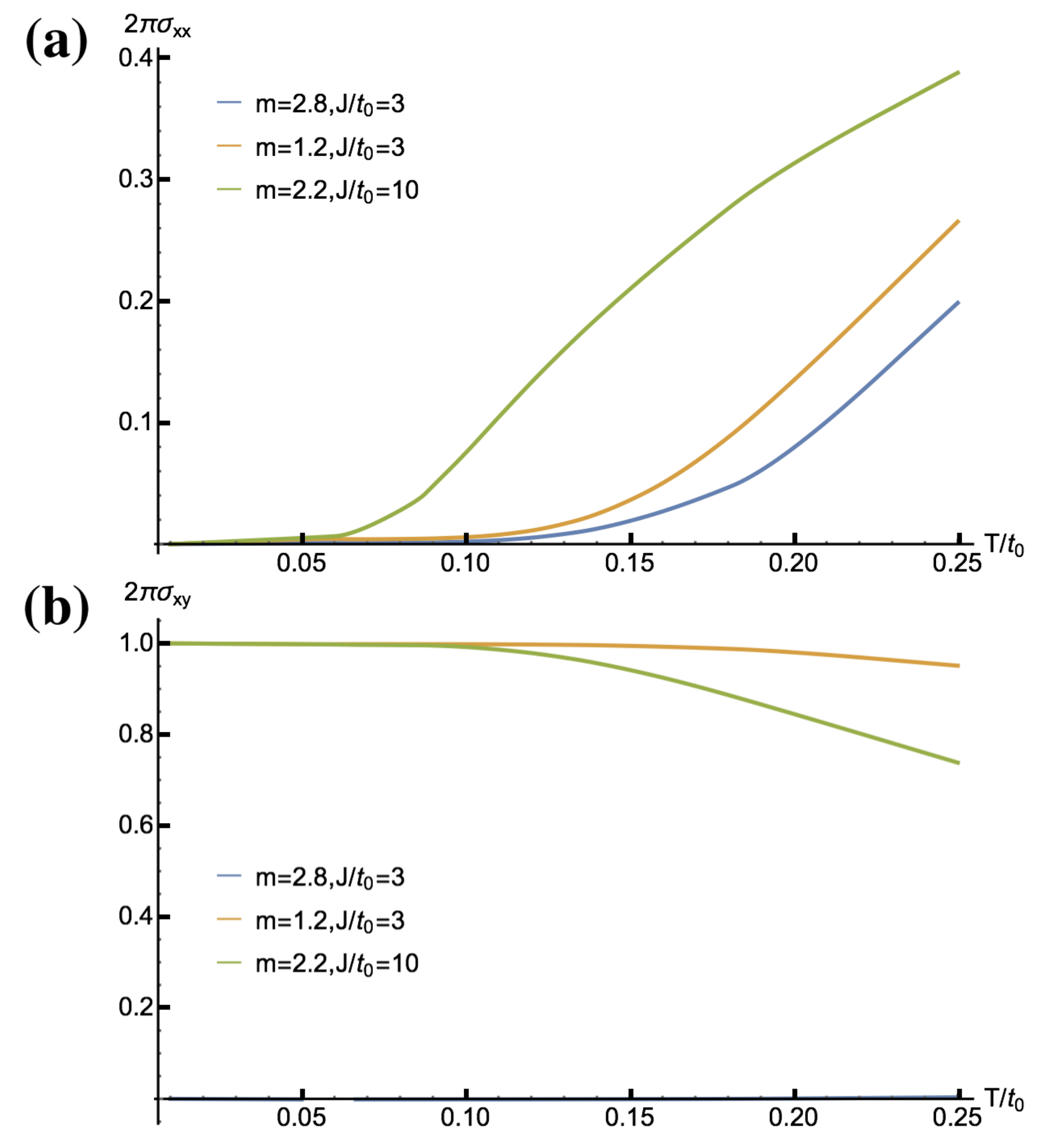}
\caption{The longitudinal (a) and Hall conductance (b) as a function of temperature at different interaction $J$ with the same vaue of $|m_\text{eff}-2|$.}
\label{conductance}
\end{figure}

\textit{Conductance.} At zero-temperature, it can be shown that the Hall conductance $\sigma_{xy}$ obeys $2\pi\sigma_{xy}=\mathcal{C}$ \cite{Hall,WZ1,WZ2}. Here, we focus on the conductance at finite temperature to see how stable the topological phases here against temperature. By implementing the Keldysh formula combined with large-N expansion, we can explore how the $U(1)$ phase response to the external electric field, and eventually we can obtain both the longitudinal and the Hall conductance directly from the two-point Green's function obtained above \cite{supple}. Here we directly show the numerical results in Fig. \ref{conductance}. 

Since the phase boundary is determined by $m_\text{eff}=\pm 2$, here we choose three points in the phase diagram that has the same values of $|m_\text{eff}-2|$, i.e. nearly equal distance away from the phase boundary. But the interaction strength of these three cases are different. We can clearly see that for larger interaction strength (the case with $m=2.2$ and $J=10t_0$), it demands lower temperature in order that the longitudinal conductance increases significantly above zero and the Hall conductance drops below quantized value. That means that in this model, although increasing interaction always drives the system into a topological nontrivial regime, the topological phase with strong interaction is less stable against temperature.    

\textit{Summary.} In summary, we construct an exact solvable interaction model to show an interaction driven quantum Hall plateau transition. We show that interaction plays two roles in this model. On one hand, it normalizes the band parameter and drives a transition from topological trivial phase to a nontrivial phase. By calculating the single particle spectral function, we show clearly that a single Dirac fermion emerges at the critical line. On the other hand, a very large interaction will eventually make the gap smaller because the SYK interaction alone favors a gapless non-Fermi liquid phase. By calculating the conductance at finite temperature, we find that the topological phase at larger interaction is more sensitive to temperature. Though the model itself is hard to realize, the physics illustrated here can shed light on other related physical systems, for instance, recent experiments have realized topological Haldane model with cold atomic gases \cite{Haldane exp,prop ja,prop hui} where the on-site repulsive interaction is tunable, and the interaction effect in the Haldane-Hubbard model is a subject of considerable research interests \cite{pre1,pre2,pre3,Huitao,para,lat1,lat2,lat3,lat4,lat5,lat6,lat7}. Our model can be viewed as a special (disordered interaction) large-N version of this model. Moreover, our model can also be extended to other topological phases, such as three-dimensional topological insulator, with the idea of granule construction based on SYK interaction.    
 
{\it Acknowledgment.} We thank Hong Yao, Zhong Wang, Cenke Xu, Chao-Ming Jian and Shao-Kai Jian for helpful discussions. This work is supported by MOST under Grant No. 2016YFA0301600 and NSFC Grant No. 11734010.

\end{document}


\title{Supplementary Material for ``Topological Sachdev-Ye-Kitaev Model"}
\author{Pengfei Zhang}
\affiliation{Institute for Advanced Study, Tsinghua University, Beijing, 100084, China}
\author{Hui Zhai}
\affiliation{Institute for Advanced Study, Tsinghua University, Beijing, 100084, China}
\affiliation{Collaborative Innovation Center of Quantum Matter, Beijing, 100084, China}

	\date{\today}
	\maketitle

	\section{Self-consistent equation in the real time.}
\begin{figure}[t]
	\centering
	\includegraphics[width=0.3\textwidth]{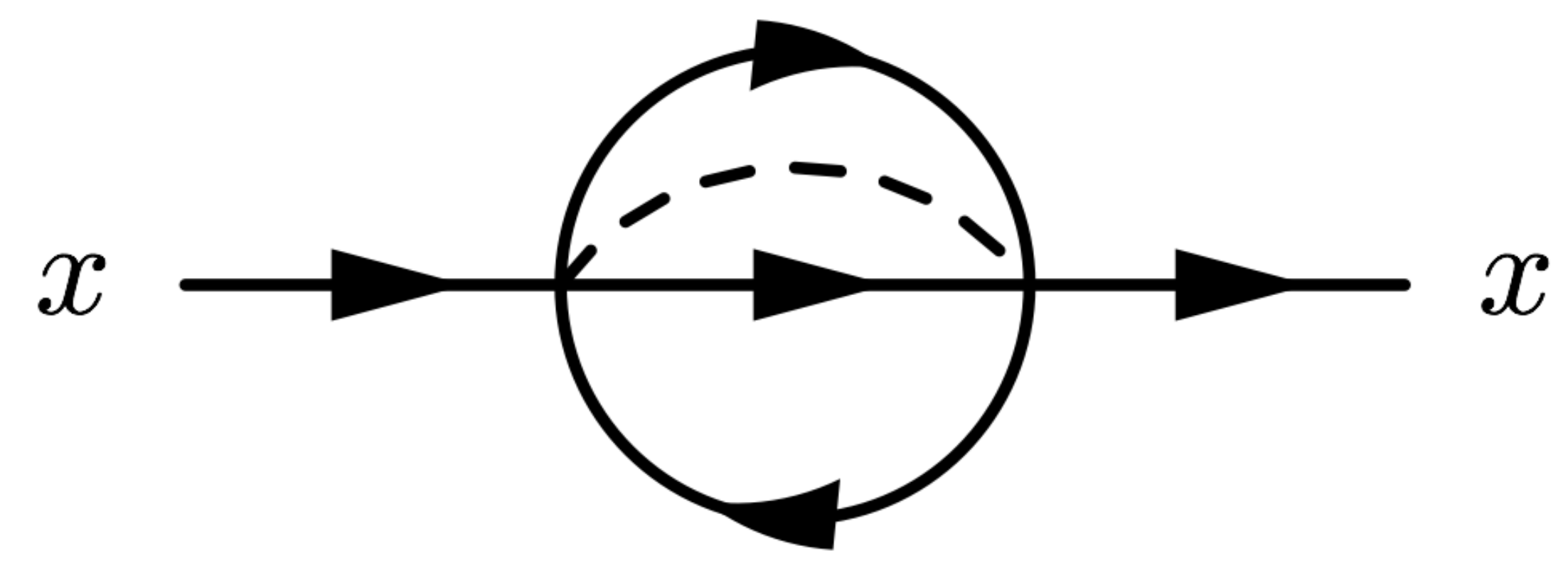}
	\caption{The self-energy diagram at the leading order of  the large $N$ expansion for the Green's function of fermions. The dashed line means the disorder average and the solid line is the full Green's function of fermions.}
	\label{melon}
\end{figure}

    In this section, we would like to present the self-consistent equaton used in our numerics. The Hamiltonian is given by the Eq. (5) of the main text. Following the standard large-N analysis for the SYK model \cite{Comments}, the only non-vaninshing diagram for the self-energy is shown in FIG. \ref{melon}. The Schwinger-Dyson equation in the imaginary time is given by:
\begin{align}
G^{-1}(i \omega_n,\mathbf p)&=-i\omega_n+h(\mathbf p)-\Sigma(i\omega_n),\label{sc1}\\ 
\Sigma(\tau)&=\begin{pmatrix}
\Sigma^{\uparrow}(\tau)&0\\
0&\Sigma^{\downarrow}(\tau)
\end{pmatrix},\label{sc2}
\end{align}
with
\begin{align}
\Sigma^{\sigma}(\tau)=J^2(G^{\sigma\sigma}(\tau))^2G^{\sigma\sigma}(-\tau).
\end{align}
The propagator $G$ here is defined as $$\left(G(\tau,\mathbf x)\right)^{\sigma\sigma'}\delta_{\lambda \lambda'}=\left<\mathcal T_\tau c_{i+\mathbf x,\sigma \lambda}(\tau)c_{i,\sigma'\lambda'}(0) \right>,$$ and we have used a simplified notion $G_\tau(\tau)\equiv G(\tau,\mathbf 0)$. Following similar derivations as in Ref.\cite{Comments,Altman,our,DPSYK,Balent} by doing an analytical continuation, we obtain the self-consistent equation for the retarded Green's function $G_{R}(\mathbf x,t)$ in the real time as
\begin{align}
G_R^{-1}(p,\omega)&=\omega-h(p)-\Sigma_R(\omega),\\ 
\Sigma_R(\omega)&=\begin{pmatrix}
\Sigma_R^\uparrow(\omega)&0\\
0&\Sigma_R^\downarrow(\omega)
\end{pmatrix},
\end{align}
where
\begin{align}
\Sigma_R^\sigma(\omega)=-iJ^2\int_0^\infty dt e^{i\omega t}((n_1^\sigma(t))^2n_2^\sigma(t)+(n_3^\sigma(t))^2n_4^\sigma(t)),\label{num1}
\end{align} 
and 
\begin{align}
&n_1^\sigma(t)=\int A^{\sigma}(\omega)n_F(-\omega)e^{-i\omega t}=n_4^\sigma(t)^*,\\
&n_2^\sigma(t)=\int A^{\sigma}(\omega)n_F(\omega)e^{i\omega t}=n_3^\sigma(t)^*.\\
&A^{\sigma}(\omega)=-\frac{1}{\pi}\text{Im}\int\frac{d^2p}{(2\pi)^2}G_R^{\sigma\sigma}(p,\omega).
\end{align}
Here $A^\sigma$ is the spectral function of fermions. This set of equations can be directly solved numerically by iteration. To get correct zero-temperature result, one should be careful about the discretization of integrals, which will be discussed in the next section.

	\section{Finite size scaling.}
When solving the self-consistent equation in the real time at $T=0$, the integration over time $t$ and frequency $\omega$ are approximated by a summation over finite number of points. This sets a cutoff which serves as an effective temperature. To obtain physical results at the zero temperature, we need to perform a finite size scaling to send the discretization $L$ to infinity. In this section we explain the details of this procedure.

In the numerics we take constant hopping $t_0$ to be the unit and the cutoff of the integration over frequency $\omega$ (time $t$) is given by $20 t_0$ (20/$t_0$). We check that the result is robust against the change of this cutoff. The most sensitive parameter is the discretization $L$ when we fix the cutoff. Instead of a uniform discretization, to increase the accuracy, we implement the Gauss-Kronrod rule to approximate the integration to the $L$-th order and perform a finite size scaling in terms of $L$.

\begin{figure}[t]
	\centering
	\includegraphics[width=1\textwidth]{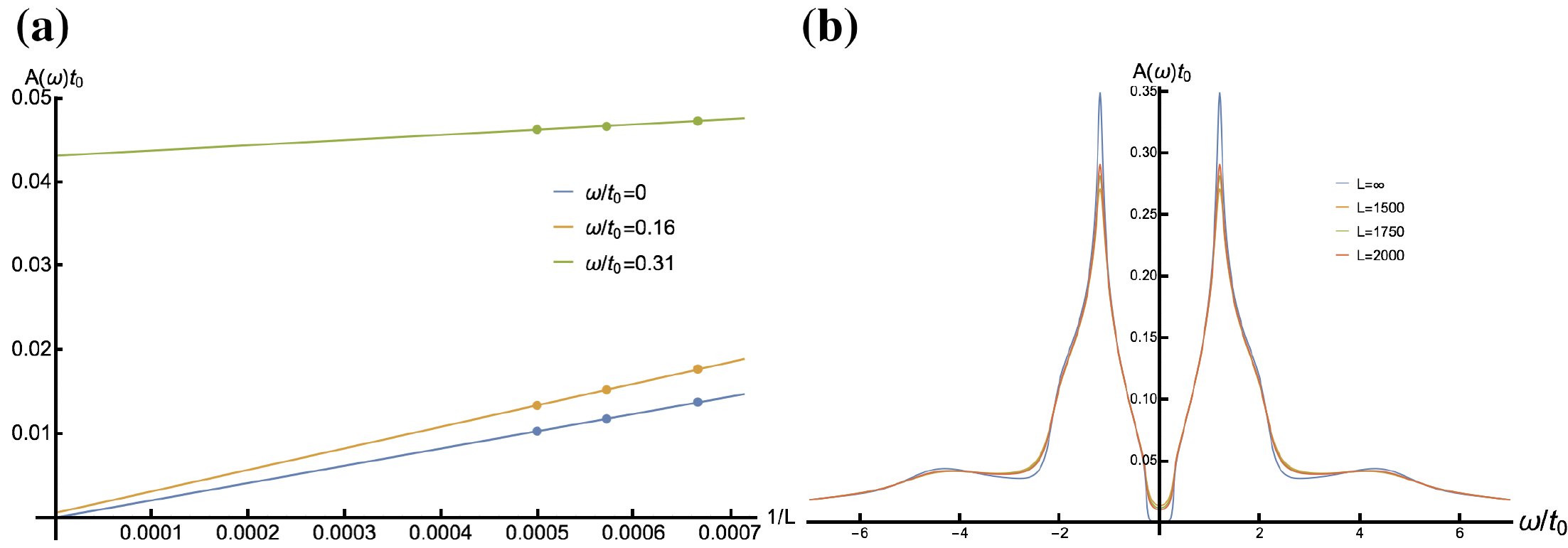}
	\caption{The numerical results for finite size scaling. (a). Spectral function at some frequency as a function of $1/L$ and the fitting to a straight line. (b). Spectral function for different $1/L$.}
	\label{fssp}
\end{figure}

A typical result is shown in FIG. \ref{fssp}. We set $J/t_0=8$ and $m=2.5$. After a finite size scaling, it is clear from Fig. \ref{fssp}(a) that for sufficiently small $\omega$, $A(\omega)$ always scales to zero and the system is gapped. From Fig. \ref{fssp}(b) the peak also becomes sharper which indicates the effective temperature is lowered. We have also checked the sum rule of the spectral function that $\int d\omega A(\omega)=1$ is always satisfied with an error of $10^{-4}$ order.

\section{symmetry and its constrain to the self-energy.}
In this section we show that because of the $\mathcal P$ symmetry defined as the Eq. (11) of the main text, there is a relation $\Sigma^\uparrow(0)=-\Sigma^\downarrow(0)$ when the self-energy at $\omega=0$ is real. Recalling the definition of the $\mathcal P$ symmetry: 
\begin{align}
\mathcal{P}c_{\mathbf p\sigma\lambda}\mathcal{P}^{-1}=(\sigma_y)_{\sigma\sigma^\prime}c^{\dagger}_{\mathbf p\sigma^\prime\lambda}, \label{P-symmetry}
\end{align}
now we study the constraint that it can impose on the Green's function:
\begin{align}
G(\omega=0,\mathbf p)=&\int d\tau\left<c_{\mathbf p,\lambda}(\tau)c^\dagger_{\mathbf p,\lambda}(0)\right>\\=&\int d\tau\left<\mathcal P\mathcal{P}^{-1}c_{\mathbf p,\lambda}(\tau)\mathcal P\mathcal{P}^{-1}c^\dagger_{\mathbf p,\lambda}(0)\mathcal P\mathcal{P}^{-1}\right>\\=&\int d\tau\ \sigma_y\left<c^\dagger_{\mathbf p,\lambda}(\tau)c_{\mathbf p,\lambda}(0)\right>(-\sigma_y)\\=&\sigma_y G(\omega=0,\mathbf p)\sigma_y
\end{align}
Here we have used the matrix representation by keeping the spin index implicit and we assume the ground state do not break the $\mathcal{P}$ symmetry. By taking the inverse of above identity, one obtains 
\begin{align}
G^{-1}(\omega=0,\mathbf p)=\sigma_y G^{-1}(\omega=0,\mathbf p)\sigma_y.
\end{align}
Knowing that $G^{-1}(\omega=0,\mathbf p)=h(\mathbf p)-\Sigma(0)$ and when the self energy is purely real, one reaches $\Sigma(0)=\sigma_y \Sigma(0)\sigma_y$ and thus $\Sigma(0)\propto \sigma_z$. Consequently, $\Sigma^\uparrow(0)=-\Sigma^\downarrow(0)$.

\section{Understanding the phase boundary: a toy model for weak interaction.}
In this section we aim at understanding the reason for why SYK interaction favors a topological nontrivial phase by a simple toy model. We approximate the original model by a simplified two-sites complex SYK model with single-particle mixing as
	\begin{align}
	H=t_0\sum\limits_{\sigma\sigma^\prime}c^\dagger_{\sigma\lambda}(m\sigma_z+ \delta \sigma_x)_{\sigma \sigma'}c_{\sigma'\lambda}+\frac{1}{4}\sum\limits_\sigma\sum\limits_{\lambda_1\lambda_2\lambda_3\lambda_4}J_{\lambda_1\lambda_2\lambda_3\lambda_4}^\sigma c^\dagger_{\sigma\lambda_1} c^\dagger_{\sigma\lambda_2} c_{\sigma\lambda_3} c_{\sigma\lambda_4},
	\end{align}
 where $J_{\lambda_1\lambda_2\lambda_3\lambda_4}^\sigma$ satisfies the Gaussian distribution as the on-site random interaction in the original model. This toy model mimics the original model when $m\gg 1$ and if we could neglect the momentum dependence of the mixing between spins. 
 Without random interaction, the retarded Green's function is given by:
	\begin{align}
	G_0(\omega)=\frac{\omega+m t_0\sigma_z+\delta t_0 \sigma_x}{\omega^2-t_0^2m^2-t_0^2\delta^2}=\frac{\omega+mt_0\sigma_z+\delta t_0\sigma_x}{\omega^2-M^2t_0^2}.
	\end{align} 
	Here we define $M=\sqrt{m^2+\delta^2}$. Focusing on fermions with spin $\uparrow$, without interaction, the spectral function is given by:
	\begin{align}
	A^\uparrow(\omega)=-\frac{1}{\pi}\text{Im}G_0^{\uparrow\uparrow}(\omega)=\frac{M+m}{2M}\delta(\omega-Mt_0)+\frac{M-m}{2M}\delta(\omega+Mt_0).
	\end{align}
	As given in Eq. \eqref{num1}, the self-energy for the retarded Green's function is given by:
	\begin{align}
	\Sigma(\omega)=-iJ^2\int_0^\infty dt\left(n_1(t)^2n_2(t)+n_3(t)^2n_4(t)\right)e^{i\omega t}.
	\end{align}
	Here $n_i(t)$ is given by:
	\begin{align}
	n_1(t)&=\int A(\omega)n_F(-\omega)e^{-i\omega t}d\omega =\frac{M+m}{2M}e^{-i M t_0 t}=n_4(t)^*,\\
    n_2(t)&=\int A(\omega)n_F(\omega)e^{i\omega t}d\omega =\frac{M-m}{2M}e^{-i M t_0t}=n_3(t)^*.
	\end{align}
	As a result one has:
 \begin{align}
	m_{\text{eff}}=m-\frac{m J^2 \delta^2}{12t_0^2(m^2+\delta^2)^2}.
 \end{align}
 Since random interaction shifts $|m|$ to be smaller, this effect stabilizes the topological phase. Physically, one could understand this by considering a single particle excitation with spin $\uparrow$. This single excitation, with energy $t_0|m|$, can be scattered to a state with two particles with spin $\uparrow$ and one hole with spin $\downarrow$, whose energy is $3t_0|m|$. As a result, the energy of a single excitation, $|m|$, is lowered.

\section{Calculation of conductance.}
In this section we explain our method for calculating the conductance. We use the Keldysh contour to formulate the field theory directly in the real time to avoid the analytical continuation \cite{Kamenev}. In the standard Keldysh approach there are two time contours $+/-$ and thus two copies of fields $c_{i,+}$ and $c_{i,-}$ (we drop flavor and spin indexes for simplicity). The phase fluctuation is introduced by $c_{i,\pm}\rightarrow \exp(-i\phi_{i,\pm})c_{i,\pm}$ with an assumption that $\phi_i$ is a smooth fluctuation approximated by $\phi_\pm(\mathbf x)$ \cite{Balent,DPSYK}. We take the convention \cite{Kamenev}: 
\begin{align}
&c_1=\frac{1}{\sqrt{2}}(c_++c_-),\ \ \ \ c_2=\frac{1}{\sqrt{2}}(c_+-c_-),\\
&\overline{c}_1=\frac{1}{\sqrt{2}}(\overline{c}_+-\overline{c}_-),\ \ \ \ \overline{c}_2=\frac{1}{\sqrt{2}}(\overline{c}_++\overline{c}_-),\\
&\phi_{\text{cl}/q}=\frac{1}{2}(\phi_{+}\pm\phi_{-}),\ \ A_{\alpha,\text{cl}/q}=\frac{1}{2}(A_{\alpha,+}\pm A_{\alpha,-}).
\end{align}
Here $A_{i,\pm}$ are gauge fields introduced by minimal coupling to extract the current-current correlation function. In the low-energy limit, the coupling is given by:
\begin{align}
S_K[A]=&A_{\alpha,+}c_+^\dagger(\partial_\alpha h)c_++\frac{1}{2}A_{\alpha,+}A_{\beta,+}c_+^\dagger(\partial_\alpha\partial_\beta h)c_+\notag\\&-A_{\alpha,-}c_-^\dagger(\partial_\alpha h)c_--\frac{1}{2}A_{\alpha,-}A_{\beta,-}c_-^\dagger(\partial_\alpha\partial_\beta h)c_-.
\end{align}
Here the partial derivative is for the corresponding component of momentums and we leave out the label of momentum for simplicity. Here we start with the standard Green's function in Keldysh formalism: $$\left<c_{i+\mathbf x,\alpha}(t)\overline{c}_{i,\beta}(0)\right>=\begin{pmatrix}
G_R(\mathbf x,t)&G_K(\mathbf x,t)\\
0&G_A(\mathbf x,t)
\end{pmatrix},$$
where $G_R$/$G_A$/$G_K$ is the retarded/advanced/Keldysh Green's function for fermions. We have $G_K(\mathbf p,\omega)=(1-2n_F(\omega))(G_R(\mathbf p,\omega)-G_A(\mathbf p,\omega))$ in thermal equilibrium which is called the fluctuation-dissipation theorem. We have omitted the spin index and all the correlation function are matrix. The current and the retarded Green's function for current operators are then given by:
\begin{align}
J_{\alpha,\text{cl}}=-\frac{i}{2}\frac{\partial \ln Z}{\partial A_{\alpha,q}},\ \ \ \ 
\Pi_{R,\alpha\beta}=\frac{\partial J_{\alpha,\text{cl}}}{\partial A_{\beta,\text{cl}}}=-\frac{i}{2}\frac{\partial^2 \ln Z}{\partial A_{\beta,\text{cl}} A_{\alpha,\text{q}}}.
\end{align} 
The longitudinal conductance and the Hall conductance are then given by:
\begin{align}
\sigma_{xx}=&\lim_{\omega\rightarrow0}\frac{1}{i\omega}\Pi_{R,xx}(\mathbf{0},\omega).\\
\sigma_{xy}=&\lim_{\omega\rightarrow0}\frac{1}{2i\omega}(\Pi_{R,xy}(\mathbf{0},\omega)-\Pi_{R,yx}(\mathbf{0},\omega)).
\end{align}

As any well-defined large-N model, after integrating out the fermions, the effective action is proporational to $N$: $S_{\text{eff}}\left[\phi\right]=NS'_{\text{eff}}\left[\phi\right]$, because of which the Green's function of $\phi$ is suppressed by $1/N$ \cite{Coleman}. As a result, only the Gaussian fluctuation of $\phi$ is taken into account in the leading order of large-$N$ \cite{Balent,DPSYK}, and therefore it only requires the information of two-point correlation function of fermions computed in the previous sections. Following similar procedures as in Ref. \cite{DPSYK}, we could derive 
\begin{align}
iS'_{\text{eff}}\left[\phi,A_{\alpha}\right]=&(D_d\delta^{\alpha\beta}+D_{pp}^{\alpha\beta})(\partial_\alpha\phi_q+A_{\alpha,q})(\partial_\beta\phi_{\text{cl}}+A_{\beta,\text{cl}})+D_{\omega\omega}\partial_t\phi_{\text{cl}}\partial_t\phi_{q}\notag\\+&D_{\omega p}^\alpha (\partial_\alpha\phi_{\text{cl}}+A_{\alpha,\text{cl}})\partial_t\phi_{q}+D_{\omega p}'^\alpha \partial_t\phi_{\text{cl}}(\partial_\alpha\phi_{q}+A_{\alpha,q}),\\
D^{\alpha\beta}_{pp}(\omega)=&-\int\frac{dq_0d^2q}{(2\pi)^3}(\text{tr}\left[\frac{\partial h}{\partial q_\alpha}G_R(q_0+\omega,\mathbf{q})\frac{\partial h}{\partial q_\beta}G_K(q_0,\mathbf{q})\right]+\text{tr}\left[\frac{\partial h}{\partial q_\alpha}G_K(q_0+\omega,\mathbf{q})\frac{\partial h}{\partial q_\beta}G_A(q_0,\mathbf{q})\right]),\\
D_{\omega\omega}(\omega)=&-\int\frac{dq_0d^2q}{(2\pi)^3}(\text{tr}\left[G_R(q_0+\omega,\mathbf{q})G_K(q_0,\mathbf{q})\right]+\text{tr}\left[G_K(q_0+\omega,\mathbf{q})G_A(q_0,\mathbf{q})\right]),\\
D^\alpha_{\omega p}(\omega)=&-\int\frac{dq_0d^2q}{(2\pi)^3}(\text{tr}\left[\frac{\partial h}{\partial q_\alpha}G_R(q_0+\omega,\mathbf{q})G_K(q_0,\mathbf{q})\right]+\text{tr}\left[\frac{\partial h}{\partial q_\alpha}G_K(q_0+\omega,\mathbf{q})G_A(q_0,\mathbf{q})\right]),
\\D'^\alpha_{\omega p}(\omega)=&-\int\frac{dq_0d^2q}{(2\pi)^3}(\text{tr}\left[G_R(q_0+\omega,\mathbf{q})\frac{\partial h}{\partial q_\alpha}G_K(q_0,\mathbf{q})\right]+\text{tr}\left[G_K(q_0+\omega,\mathbf{q})\frac{\partial h}{\partial q_\alpha}G_A(q_0,\mathbf{q})\right]).
\end{align}
Here we keep the label of freqency implicit and only show the retarded part of the action, and $D_d=-D_{pp}^{\alpha\alpha}(0)$ is the contribution from the diamagnetic term. In the zero momentum limit, one could show $D_{\omega p}^\alpha=D_{\omega p}'^\alpha=0$, we find now:
\begin{align}
\sigma_{xx}=&-\lim_{\omega\rightarrow0}\frac{1}{2\omega}(D^{xx}_{pp}(\omega)-D^{xx}_{pp}(0)),\label{long con}\\
\sigma_{xy}=&-\lim_{\omega\rightarrow0}\frac{1}{4\omega}(D^{yx}_{pp}(\omega)-D^{xy}_{pp}(\omega)).\label{Hall con}
\end{align}
Eq.\eqref{Hall con} is consistent with the exact result derived in \cite{Hall} by using the Wald identity.